\documentclass[aps,prd,preprint,superscriptaddress,tightenlines,nofootinbib,floatfix]{revtex4}
\usepackage{graphicx}
\usepackage{dcolumn}
\usepackage{bm}
\usepackage{verbatim}

\newcommand{\cbarc}{$c \bar{c}$}

\newcommand{\Uos}{\Upsilon(1\text{S})}
\newcommand{\Utwos}{\Upsilon(2\text{S})}

\newcommand{\Jpsi}{$J/\psi$}
\newcommand{\Ups}{$\Upsilon$}
\newcommand{\etap}{$\eta^{\prime}$}

\newcommand{\etab}{$\eta_{b}$}
\newcommand{\pizero}{$\pi^{0}$}

\newcommand{\pbi}{\,\rm pb^{-1}}
\newcommand{\fbi}{\,\rm fb^{-1}}
\newcommand{\chisq} {$\chi^{2}$}

\newcommand{\diphoton}{$\gamma \gamma$}

\newcommand{\dedx}{$dE/dx$}

\newcommand{\chisqFourMom}{$\chi^{2}_{\text{P}4}$}
\newcommand{\EtaMassFitChisq} {$\chi^{2}_{\eta}$}

\newcommand{\chisqTotal}{$\chi^{2}_{\mathrm{Total}}$}

\newcommand{\chisqpizero}{$\chi^2_{\pi^0}$}

\newcommand{\Nups}{$N_{\Uos}$}

\newcommand{\sensitvt}{\epsilon_{i}\cdot\mathcal{B}_{\text{P},i}\cdot N_{\Uos}}

\newcommand{\EtaGG}{\eta \to \gamma \gamma}
\newcommand{\EtaPMZ}{\eta \to \pi^{+} \pi^{-} \pi^{0}}
\newcommand{\EtaThreePZ}{\eta \to \pi^{0} \pi^{0} \pi^{0}}

\newcommand{\EtapGR}{$\eta^{\prime} \to \gamma \rho^0$}
\newcommand{\ModeEta}{$\Uos \to \gamma \eta$}
\newcommand{\ModeEtap}{$\Uos \to \gamma \eta^{\prime}$}

\newcommand{\ModeEtaGG}{$\Uos \to \gamma \eta; \EtaGG$}
\newcommand{\ModeEtaPMZ}{ \Uos \to \gamma \eta; \EtaPMZ }
\newcommand{\ModeEtaThreePZ}{ \Uos \to \gamma \eta; \EtaThreePZ }
\newcommand{\EtapEtaGG}{$\eta^{\prime}; \EtaGG$}
\newcommand{\EtapEtaPMZ}{$\eta^{\prime}; \EtaPMZ$}
\newcommand{\EtapEtaThreePZ}{$\eta^{\prime}; \EtaThreePZ$}
\newcommand{\ModeEtapGG}{$\Uos \to \gamma \eta^{\prime}; \EtaGG$} 
\newcommand{\ModeEtapPMZ}{$\Uos \to \gamma \eta^{\prime}; \EtaPMZ$} 
\newcommand{\ModeEtapThreePZ}{$\Uos \to \gamma \eta^{\prime};\EtaThreePZ$}
\newcommand{\ModeEtapGR}{$\Uos \to \gamma \eta^{\prime};$ \EtapGR } 
\newcommand{\qedOmega}{$e^{+} e^{-} \to \gamma \omega$}
\newcommand{\qedPhi}{$e^{+} e^{-} \to \gamma \phi$}

\newcommand{\qedThreePhoton}{$e^{+} e^{-} \to \gamma \gamma \gamma$}

\newcommand{\phiKsKl} {$\phi \to K_{S} K_{L}$}

\newcommand{\fTwo} {$f_2(1270)$}
\newcommand{\UpsToGammaFTwo} {$\Uos \to \gamma f_{2}(1270)$}
 
\newcommand{\BrModeEta}{$\mathcal{B}$(\ModeEta)}
\newcommand{\BrModeEtap}{$\mathcal{B}$(\ModeEtap)}
\newcommand{\gev}{\ \rm GeV}
\newcommand{\mev}{\ \rm MeV}

\newcommand{\mom}{\,\rm GeV/$c$}
\newcommand{\mass}{\,\rm GeV/$c^2$}
\newcommand{\miss}{\,\rm MeV/$c^2$}
\newcommand{\etal}{{\it et al.,}}

\newcommand{\br}[1]{$#1 \times 10^{-6}$}

\newcommand{\ulbr}[1]{$< #1 \times 10^{-6}$}

\begin{document}
\preprint{CLNS 07/1995}       
\preprint{CLEO 07-05}         

\title{Search for Radiative Decays of $\Uos$ into $\eta$ and \etap}

\author{S.~B.~Athar}
\author{R.~Patel}
\author{V.~Potlia}
\author{H.~Stoeck}
\author{J.~Yelton}
\affiliation{University of Florida, Gainesville, Florida 32611}
\author{P.~Rubin}
\affiliation{George Mason University, Fairfax, Virginia 22030}
\author{C.~Cawlfield}
\author{B.~I.~Eisenstein}
\author{I.~Karliner}
\author{D.~Kim}
\author{N.~Lowrey}
\author{P.~Naik}
\author{C.~Sedlack}
\author{M.~Selen}
\author{E.~J.~White}
\author{J.~Wiss}
\affiliation{University of Illinois, Urbana-Champaign, Illinois 61801}
% ...
\collaboration{CLEO Collaboration} 
\noaffiliation

\noaffiliation
\date{April 19, 2007}

\begin{abstract} 

We report on a search for the radiative decay of $\Uos$ to the
pseudoscalar mesons $\eta$ and \etap\ in $(21.2\pm0.2)\times10^{6}$ $\Uos$
decays collected with the CLEO~III detector at the Cornell
Electron Storage Ring (CESR). The $\eta$ meson was
reconstructed in the three modes $\EtaGG$, $\EtaPMZ$ or $\EtaThreePZ$.
The \etap\ meson was reconstructed in the mode
$\eta^{\prime} \to \pi^{+} \pi^{-} \eta$ with $\eta$ decaying through any
of the above three modes, and also
$\eta^{\prime} \to \gamma \rho^0$, where $\rho^0 \to \pi^+ \pi^-$.

Five out of the seven sub-modes are found to be virtually background-free.
In four of them we find no signal candidates and in one 
($\Uos\to\gamma\eta^{\prime},~\eta^{\prime}\to\pi^+\pi^-\eta,~\EtaPMZ$)

there are two good signal candidates, which is insufficient evidence
to claim a signal.

The other two sub-modes ($\EtaGG$ and $\eta^{\prime} \to \gamma \rho^0$)
are background limited, and show no excess of events in their signal
regions. We combine the results from different channels and obtain
upper limits at the 90\% C.L. which are 
$\mathcal{B}(\Uos \to \gamma \eta) < 1.0 \times 10^{-6}$ and
$\mathcal{B}(\Uos \to \gamma \eta^{\prime}) < 1.9 \times 10^{-6}$. 
Our limits are an order of magnitude tighter than the previous
ones and below the predictions made by some theoretical models.

\end{abstract}

\maketitle

\section{Introduction}
The hadronic decays of heavy quarkonia below the threshold for heavy 
meson pair production are understood to proceed predominantly via
three intermediate gluons. One of the gluons can be replaced by a 
photon with a penalty of order the ratio of coupling constants, 
$\alpha / \alpha_s$. Such exclusive radiative decays of the heavy vector
mesons \Jpsi\ and \Ups\ have been the subject of many experimental and
theoretical studies. For the experimenter, the final states from
radiative decays are relatively easy to identify as they have a high
energy photon, a low multiplicity of other particles, and low
background. Theoretically, the radiative decays of heavy quarkonia
into a single light hadron provide a particularly clean environment to 
study the conversion of gluons into hadrons, and thus their study is a
direct test of QCD. \ModeEtap\ is one such candidate channel. This
decay channel has been observed to be produced in the \Jpsi\
charmonium system (the $1^3\text{S}_1$ state of \cbarc) with
$\mathcal{B}(J/\psi\to\gamma \eta^{\prime}) = (4.71\pm0.27)\times10^{-3}$~\cite{PDG}.  
Naive scaling predicts that decay rates for radiative $\Uos$ decays
are suppressed by the factor 
$(q_{b}m_{c}/q_{c}m_{b})^{2}$ $\approx 1/40$ 
with respect to the corresponding \Jpsi\ radiative decays.
This factor arises because the quark-photon coupling is proportional
to the electric charge, and the quark propagator is roughly
proportional to $1/m$ for low momentum quarks. Taking into account the
total widths~\cite{PDG} of \Jpsi\ and $\Uos$, the branching fraction
of a particular $\Uos$ radiative decay mode is expected to be around
0.04 of the corresponding \Jpsi\ branching fraction. However, the CLEO 
search~\cite{CleoEtaPrimeStudy} for \ModeEtap\ in $61.3\pbi$ of data
collected with the CLEO~II detector found no signal in this mode, and
resulted in a 90\% confidence level 
upper limit of $1.6\times10^{-5}$ for the branching
fraction \ModeEtap, an order of magnitude smaller than this
expectation.

The two-body decay \UpsToGammaFTwo\ has been
observed~\cite{CleoF2_1270} in the older CLEO~II $\Uos$ analysis,
and this observation has been confirmed~\cite{LuisAnalysis,Holger}, with much
greater statistics, in CLEO~III data. 
The measurement $\mathcal{B}(\Uos\to\gamma f_2(1270)) = (10.2\pm1.0) \times10^{-5}$, 
from the combination of the two CLEO~III measurements, 
is $0.074\pm0.010$ times the corresponding \Jpsi\
decay mode, showing a deviation of roughly a factor of two from the
naive scaling estimates.
In radiative \Jpsi\ decays the
ratio of \etap\ to \fTwo\ production is $3.4\pm0.4$. If the same ratio
held in $\Uos$, the \etap\ channel would be clearly visible.
The channel \ModeEta\ has
received significant theoretical attention. 
This channel has been observed in \Jpsi\ decays~\cite{PDG}
with the branching fraction of $(9.8\pm1.0)\times 10^{-4}$, a value
smaller by a factor of 
five
than $\mathcal{B}(J/\psi\to\gamma \eta^{\prime})$. 
The previous CLEO search of $\Uos$ decays produced an upper limit of
$2.1\times10^{-5}$ at the 90\% confidence level 
for this mode~\cite{CleoEtaStudy}.

Several authors have tried to explain the lack of signals in radiative 
$\Uos$ decays into pseudoscalar mesons, using a variety of models
which produce branching fraction predictions of 
$10^{-6}\ \text{to}\ 10^{-4}$.
Employing the Vector 
Meson Dominance Model (VDM), Intemann~\cite{Intemann} predicts the
branching fractions for the heavy vector meson radiative decay into
light pseudoscalar mesons. Using  the mixing mechanism of $\eta$,
\etap\ with the as-yet-unobserved pseudoscalar resonance \etab,
Chao~\cite{KTChao} first calculated the mixing angle
$\lambda_{\eta\eta_{b}}$ in order to estimate the radiative branching
fractions. Baier and Grozin~\cite{BaierGrozin} showed that for light
vector mesons (such as \Jpsi) there might be an additional ``anomaly''
diagram that contributes significantly to the radiative decays. Noting
that VDM has no direct relation to QCD as the fundamental theory of
strong interactions, and referring to~\cite{Intemann},
Ma tries to address the problem by using factorization at tree level
with NRQCD matrix elements to describe the heavy vector meson portion
multiplied by a set of twist-2 and twist-3 gluonic distribution
amplitudes~\cite{JPMa}.

\section{Detector and Data Sample}
This study is based upon data collected by the CLEO III detector
at the Cornell Electron Storage Ring (CESR). CLEO III is
a versatile multi-purpose particle detector described fully
elsewhere~\cite{cleoiii-detector}. Centered on the $e^+e^-$ interaction
region of CESR, the inner detector consists of a silicon strip vertex
detector and a wire drift chamber measuring the momentum vectors and
the ionization energy losses (\dedx) of charged tracks based on their
trajectories in the presence of a 1.5T solenoidal magnetic field. The
silicon vertex detector and the drift chamber tracking system together
achieve a charged particle momentum resolution of 0.35\% (1\%) at
1\mom\ (5\mom)
and a fractional \dedx\ resolution of 6\% for hadrons and 5\% for electrons. 
Beyond the drift chamber is a Ring Imaging Cherenkov Detector, RICH,
which covers 80\% of the solid angle 
and is used to further identify charged particles by giving for each
mass hypothesis the fit likelihood to the measured Cherenkov radiation
pattern. After the RICH is a CsI crystal calorimeter that covers 93\%
of the solid angle, allowing both photon detection and electron
suppression. The calorimeter provides an energy 
resolution of 2.2\% (1.5\%) for 1\gev\ (5\gev) photons. Beyond the calorimeter 
is a superconducting solenoidal coil providing the magnetic field,
followed by iron flux return plates with wire chambers interspersed 
at 3, 5, and 7 hadronic interaction lengths (at normal 
incidence) to provide
muon identification.

The data sample has an integrated luminosity of $1.13\fbi$ taken at the
$\Uos$ energy $\sqrt{s} = 9.46\gev$, which corresponds to 
$N_{\Uos} = 21.2\pm0.2$ million $\Uos$ decays~\cite{CLEO-III-NUPS}. 
The efficiencies for decay chain reconstruction were obtained from
Monte 
Carlo simulated radiative events generated with the 
($1+\cos^{2}\theta$) angular distribution expected for decays 
$\Uos \to \gamma+\text{pseudoscalar}$. The Monte Carlo simulation of
the detector response was based upon GEANT~\cite{GEANT}, and
simulation  events were processed in an identical fashion to data.

\newcommand{\SigPi}{$\sigma_{\pi,i}$}
\newcommand{\SigPiSq}{$\sigma^{2}_{\pi}$}
\newcommand{\SigSumPizSq}{$\sum_{i}^{3}\sigma^{2}_{\pi,i}$}

\newcommand{\SPi}{$S_{\pi,i}$}
\newcommand{\SPiSq}{$S^{2}_{\pi}$}
\newcommand{\SSumPizSq}{$\sum_{i}^{3}S^{2}_{\pi,i}$}

\newcommand{\GGMassDiffOverSigma}{$(m_{\gamma\gamma} - m_{\pi^0})/\sigma_{\gamma\gamma}$}
\newcommand{\MassDiffOverSigma}{$((M_{reconstructed}-M_{\pi^0})/resolution)$}
\newcommand{\SDeDxDefn}{\text{S}_{dE/dx} \equiv
(dE/dx(\text{measured}) - dE/dx(\text{expected}))/\sigma_{dE/dx}}
\newcommand{\SDeDx}{\text{S}_{dE/dx}}
\newcommand{\SDeDxSq}{$\text{S}_{dE/dx}^{2}$}
\section{Event Selection and Results}\label{sec:event-selection}
In our search for \ModeEta\ and \ModeEtap, we reconstruct $\eta$
mesons in the modes $\EtaGG$, $\EtaPMZ$, and $\EtaThreePZ$; the
latter two will collectively be referred to as $\eta\to3\pi$. We
reconstruct the
\etap\ meson in the mode $\eta \pi^+ \pi^-$ with
$\eta$ decaying in any of the above modes, and in addition, 
the mode \EtapGR, where $\rho^0 \to \pi^+\pi^-$. 
From the CLEO~II studies~\cite{CleoEtaPrimeStudy,CleoEtaStudy} we
expected five out of the seven modes under investigation to be relatively
background free and so we employ minimal selection
criteria to maximize sensitivity and minimize possible systematic
biases. 
The other two, $\EtaGG$ and \EtapGR, have large branching
fractions, but also large backgrounds, and so our event selection for these
modes aims to decrease the background with a corresponding loss of
efficiency.  

Our general analysis strategy is to reconstruct the complete decay
chain ensuring that none of the constituent tracks or showers have
been used more than once, then kinematically constrain the intermediate
\pizero\ and $\eta$ meson candidates  to their nominal masses~\cite{PDG},
and finally require the event to be consistent with having the 4-momentum of the
initial $e^+e^-$ system. Multiply-reconstructed $\Uos$ candidates in an
event, a problem of varying severity from mode to mode, is dealt with 
by selecting the combination with lowest \chisqTotal, the sum of
chi-squared of the 4-momentum constraint (\chisqFourMom) and chi-squared of
all the mass-constraints involved in a particular decay chain. For
example, there are four mass-constraints involved in the decay chain
\ModeEtapThreePZ, three \pizero\ mass-constraints and one $\eta$
mass-constraint. 
The mode $\ModeEtaThreePZ$ is an exception in which we
preferred to accept the $\EtaThreePZ$ candidate having the lowest
\SPiSq\ $\equiv$ \SSumPizSq, with \SPi\ $\equiv$
\GGMassDiffOverSigma\ of the \textit{i}th \pizero candidate.  
The yield is obtained by counting the number of final state $\eta$ or
\etap\ candidates within our acceptance mass window defined as the
invariant mass region centered around the mean value and providing
98\% signal acceptance as determined from signal Monte Carlo. Whenever
possible, an event vertex is calculated using the information from
the charged tracks, and the 4-momentum of the photon candidates is then
recalculated, assuming that the showers originate from the event
vertex rather than the origin of the CLEO coordinate system. This produces
an improvement in the $\eta$ and \etap\
candidates' invariant mass resolution of roughly 10\%, leading to a slight
increase in the sensitivity of the measurement.

The CLEO~III trigger~\cite{cleoiii-trigger} relies upon two components:
(1) the tracking-based ``axial'' and ``stereo'' triggers derived from
the signals on the 16 axial layers of the drift chamber, and  the signals
registered on the chamber's 31 stereo layers, and (2) the calorimeter-based
trigger derived from the energy deposition in the CsI crystal
calorimeter. The events for the ``all neutral'' modes \ModeEtaGG\ and
$\ModeEtaThreePZ$ are collected by the calorimeter-based trigger
condition requiring two high energy back-to-back showers.
We demand that triggered events meet the
following analysis requirements: (a) a high energy calorimeter shower not
associated with a charged track, having a lateral profile consistent
with being a photon, and  having a measured energy greater than
4.0\gev\ must be present; 
(b) there must be the correct
number of pairs of oppositely charged, 
good quality tracks with usable \dedx\ information. 
The efficiency of these requirements is more than 60\% in
modes involving charged tracks and approximately 54\% and 45\% 
for cases where $\EtaGG$ and $\eta\to 3\pi^{0}$, respectively.

The photon candidates we use in forming \pizero\ and $\EtaGG$ candidates
have minimum energy depositions of 30\mev\ and 50\mev,
respectively. All photon candidates are required to be not associated
to charged tracks, and at least one of the photon candidates of 
each pair must have
a lateral profile consistent with that expected for a photon. The
photon  candidates we use in reconstructing the $\eta$ meson in the
\diphoton\ mode must be detected either in the fiducial barrel or
the fiducial endcap\footnote{The fiducial regions of the barrel and 
endcap are defined by 
$|\cos(\theta)|<0.78$ and $0.85<|\cos(\theta)|<0.95$,
respectively; the region between the barrel fiducial region and the
endcap fiducial region is not used due to its relatively poor
resolution. 
} calorimeter region only. 
These candidates are then kinematically
constrained to the nominal meson mass, the exception being \ModeEtaGG,
where no mass-constraining was done to the $\eta$ candidate, because
we examine $m_{\gamma\gamma}$ in this mode to determine our yield.

The $\eta$ candidates in the mode $\pi^+\pi^-\pi^0$ are built by first
forcing pairs of oppositely charged quality tracks to originate from a
common vertex. The \pizero\ candidate having invariant mass within
$7\sigma_{\gamma\gamma}$ is then added to complete the reconstruction
of $\EtaPMZ$ candidates. The charged tracks are required to be
consistent with being pions by adding the pion hypothesis $\SDeDxDefn$
in quadrature for two tracks and requiring the sum of \SDeDxSq\ to be
less than 16. 

In the case of $\EtaThreePZ$, the $\eta$ candidate is simply built by
adding three different \pizero\ candidates, where no constituent photon
candidate contributes more than once in a candidate $\EtaThreePZ$
reconstruction. The \pizero\ candidates are selected by requiring
$S_{\pi}<10.0$. In order to increase the efficiency in this mode, 
an exception was made to the fiducial region requirement, and
photons in the gap between the barrel and endcap fiducial regions
were allowed.

\subsection{The Decay $\Upsilon\to\gamma\eta ,\eta\to 3\pi$}
The $\Upsilon$ candidate in the mode $\gamma \eta$ is
formed by combining a high-energy photon ($E>4\gev$) with the $\eta$
candidate, requiring that this photon is not a daughter of the $\eta$
candidate. The $\Upsilon$ candidate is then subjected to the
4-momentum constraint of the initial $e^+e^-$ system. In the case of
$\eta\to3\pi$, multiply reconstructed $\Upsilon$ candidates were
restricted by selecting only one candidate. 
For $\EtaPMZ$, we select the candidate with the lowest \chisqTotal,
the sum of chi-squared of the 4-momentum constraint 
and chi-squared of the mass-constraint to the \pizero\ candidate.
For $\EtaThreePZ$, we select the candidate with the smallest \SPiSq.
The selected $\Upsilon$ candidate is further required
to satisfy the 4-momentum consistency criterion, restricting 
\chisqFourMom\ $<100$ for $\EtaPMZ$ and a less stringent cut of 200
for $\EtaThreePZ$ measurements.
In addition, we limit the number of reconstructed calorimeter showers 
for the mode $\ModeEtaThreePZ$ to minimize backgrounds such as
\qedPhi\ where \phiKsKl\ without jeopardizing the signal efficiency.

From Monte Carlo simulations, the overall reconstruction efficiencies,
$\epsilon_i$, for each channel are determined to be $(28.5\pm4.3)\%$
and $(11.8\pm1.9)\%$ for the decay chains $\Upsilon\to\gamma\eta,\EtaPMZ$ and
$\Upsilon\to\gamma\eta, \EtaThreePZ$, respectively. 
The uncertainties in the efficiency
include the Monte Carlo samples' statistical uncertainty and our
estimate of possible systematic biases, which are discussed further
in Section~\ref{sec:sys-lim}.

\begin{figure}
\includegraphics*[width=3.25in]{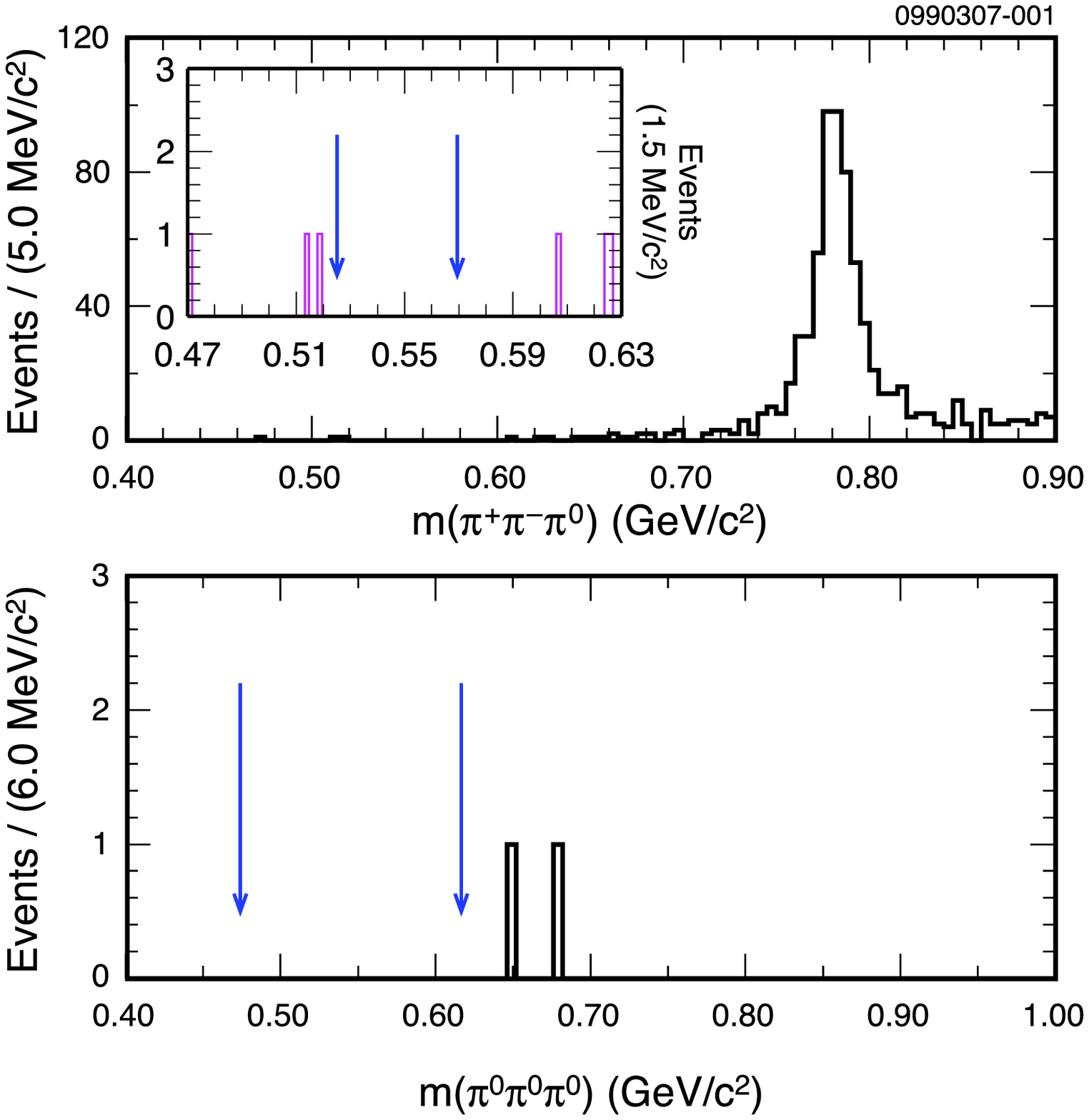}
\caption{Candidate $\EtaPMZ$ (top) and $\EtaThreePZ$ (bottom)
invariant mass distributions from $\Uos$ data.
The large number of events near 780\miss\ (top) is due to the 
abundant process \qedOmega. No events are observed in our acceptance
region, bounded by the arrows.}  
\label{fig:eta-3pi}
\end{figure}

We find no candidate events within our acceptance invariant mass
window for the search \ModeEta, $\eta\to3\pi$. The invariant mass
distributions for candidate $\EtaPMZ$ and $\EtaThreePZ$, after
imposing all the selection criteria are shown in Figure~\ref{fig:eta-3pi}.

\subsection{The Decay $\Upsilon\to\gamma\eta ,\eta\to\gamma\gamma$}
The 3-photon final state resulting from \ModeEtaGG\ is dominated
by the QED process \qedThreePhoton. Our selection criteria of loosely
reconstructing an $\EtaGG$ meson and requiring the \chisq\ of
4-momentum constraint on the $\Uos$ meson formed by adding a
hard-photon to be $<200$ are not sufficient to suppress this
background. The QED background, however has a distinct feature - the
two photons having energies 
$E_{hi}$ and $E_{lo}$ used in reconstructing the $\eta$ candidate have
a large energy asymmetry, where asymmetry is defined as
$(E_{hi}-E_{lo})/(E_{hi}+E_{lo})$.  
Real $\eta$ mesons are expected to have an approximately
uniform distribution of
asymmetry in the range (0,1). We require the asymmetry to be less than
0.8. To further discriminate between the signal and the background, we
used a neural net approach. 

The input to the neural net is a vector of six variables, namely the
measured energy and the polar angle $\theta$ of each of the three calorimeter
showers used in the reconstruction chain. The training sample is
comprised of 20,000 simulated signal and background events in equal
proportion. The simulated \qedThreePhoton\ background events have a
high-energy photon ($E>4\gev$), \diphoton\ invariant mass for the two
lower-energy photons in the range 0.4-0.7\mass, and energy asymmetry
less than 0.8.

\begin{figure}
\includegraphics*[width=3.25in]{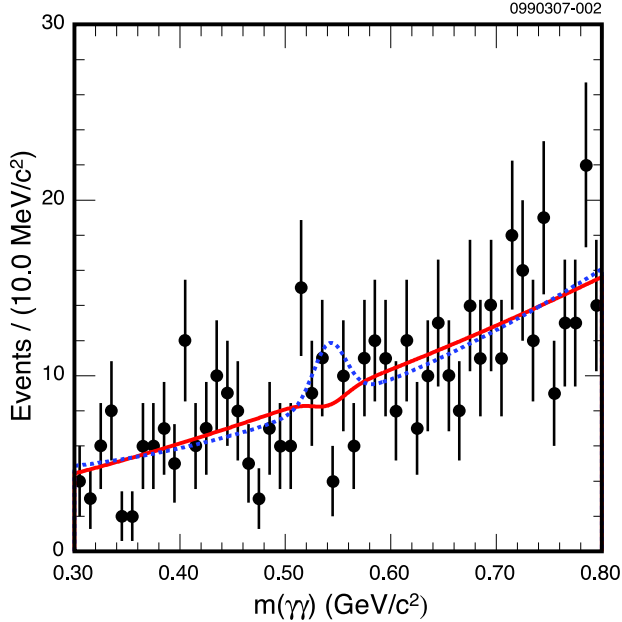}
\caption{Invariant mass distribution of \diphoton\ candidates in $\Uos$
  data for the mode \ModeEtaGG, overlaid with fits using a) floating area
  (solid red) yielding $-2.3\pm8.7$ events, and b) area fixed to 14.5
  events (dashed blue), the upper limit corresponding to 90\% C.L.}
\label{fig:eta-gg}
\end{figure}

For our final selection, we choose neural-net output with $51\%$
efficiency while rejecting $86\%$ of the background.
The combined efficiency of our
selection criteria for this mode is $(23.8\pm2.4)\%$, which includes
possible systematic biases and statistical uncertainties from the
simulation. The resulting \diphoton\ invariant mass distribution from
$\Uos$ data is fit, as shown in Figure~\ref{fig:eta-gg},
to a double Gaussian function, whose mass and widths are fixed to values
found from signal Monte Carlo data, along with a second order polynomial
background function. From this likelihood fit, we obtain $-2.3\pm8.7$
events; consistent with zero. We then perform the same likelihood fit
multiple times fixing the signal area to different values, assigning
each of the fits a probability proportional to  $e^{-{\chi^2}/2}$,
where \chisq\ is obtained from the likelihood fit.
The resulting probability distribution is normalized and numerically
integrated up to 90\% of the area to obtain the yield at 90\%
confidence level. Our limit thus obtained is 14.5 events at 90\%
confidence level. 

\subsection{The Decay  $\Upsilon\to\gamma\eta^{\prime} ,\eta^{\prime}\to\eta\pi^+\pi^-$}
Reconstruction of the decay chains \ModeEtap, where 
$\eta^{\prime} \to \eta \pi^+ \pi^-$, builds on the search \ModeEta\. 
The reconstructed $\eta$ candidate is
constrained to the nominal $\eta$ mass. The mass-constrained $\eta$
candidate is further combined with a pair of oppositely charged
quality tracks by forcing the tracks and the $\eta$ candidate to
originate from a common vertex. In reconstruction of \EtapEtaPMZ,
care is exercised to ensure that no track is used more than once in the
decay chain. The high energy photon is combined with the \etap\
candidate to build an $\Upsilon$ candidate which is further
constrained to the 4-momentum of the initial $e^+e^-$ system. 
In the reconstruction chain \EtapEtaGG, the $\Upsilon$ candidate with
the lowest sum of chi-squared to the 4-momentum constraint (\chisqFourMom)
combined with the chi-squared of the mass-constraint to the $\eta$
candidate (\EtaMassFitChisq) is accepted as
the representative $\Upsilon$ candidate in the reconstructed event.
In the modes where $\eta\to3\pi$, 
the \pizero\ mass-constraint chi-squared, \chisqpizero, also contributes to the 
\chisqTotal .

To ensure that only good quality $\eta$ candidates participate in the decay
chain, the \EtaMassFitChisq\ values of ``$\eta\to \text{all neutral}$''
candidates are required to be less than 200. Owing to the better
measurements of charged track momenta, this criterion is more
stringent (\EtaMassFitChisq $<100$) in the case of $\EtaPMZ$. The 
targeted efficiency (around 99\%) of this requirement is achieved in
all three cases.
 
The charged tracks used in reconstructing \etap\ candidates have to be
consistent with the pion hypothesis.
We again require the sum of squared $\SDeDx$ added in quadrature to
be less than 16 for both the two track and four track cases. The
efficiency of this requirement alone is around 99\%.

The selected $\Upsilon$ candidate is further required to satisfy the
4-momentum consistency criterion, restricting \chisqFourMom\ $<100$
in the $\EtaGG$ case and a less stringent value of 200 for
$\eta\to3\pi$. The overall reconstruction efficiencies of our
selection criteria as determined from signal Monte Carlo simulations are
$(35.3\pm5.2)\%$, $(24.5\pm2.2)\%$ and $(14.4\pm2.9)\%$ for $\eta$
decays to \diphoton, $\pi^+\pi^-\pi^0$ and $3\pi^0$, respectively.

\begin{figure}
\includegraphics*[width=3.44in]{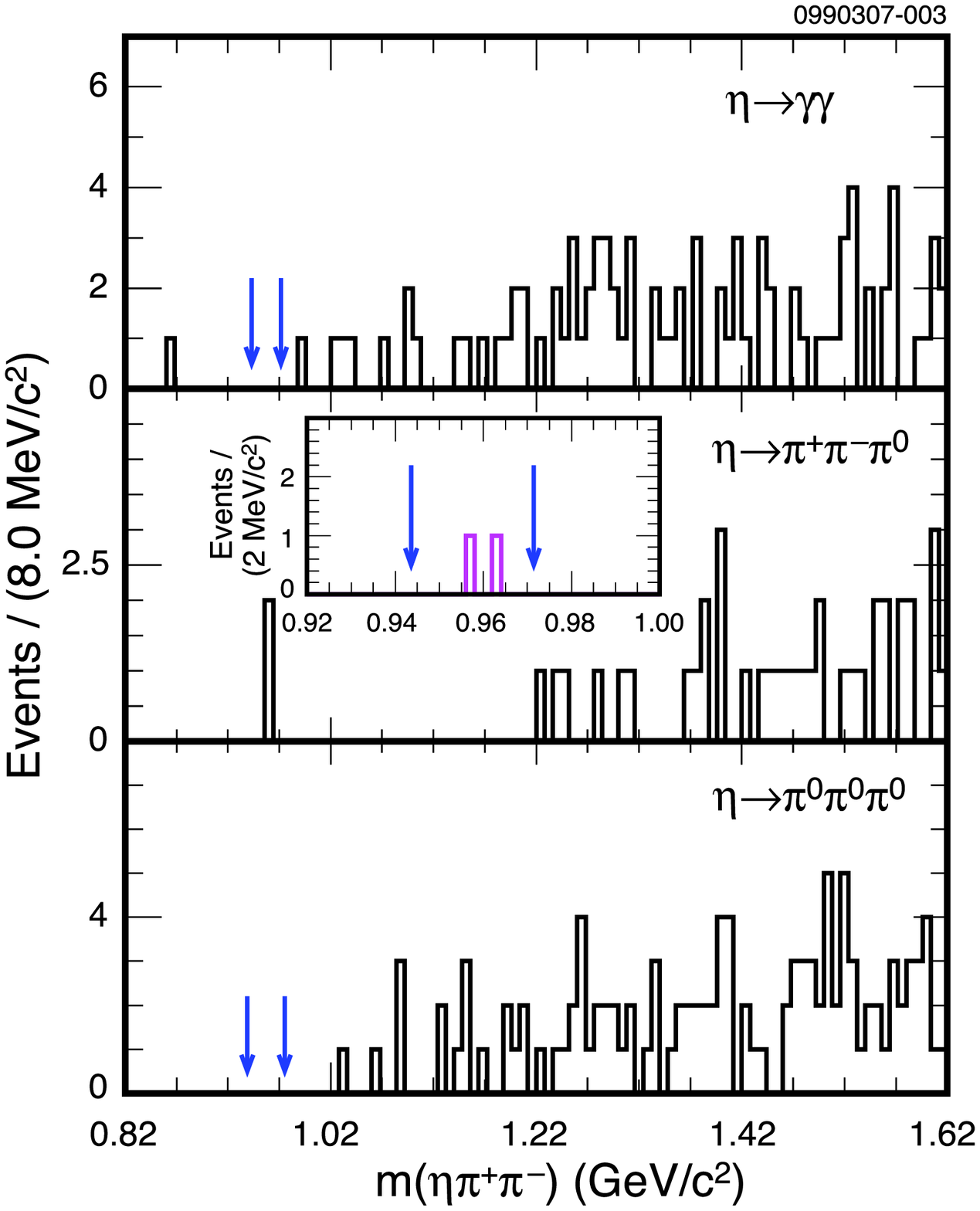}
\caption{Invariant mass distributions of $\eta\pi^+\pi^-$ candidates from $\Uos$
data. The $\eta$ candidate is constrained to the nominal $\eta$ meson
mass. No events are observed in the signal box for $\EtaGG$ (top) and
$\EtaThreePZ$ (bottom); two signal events are observed for
$\EtaPMZ$ (middle).}
\label{fig:etap-eta-pi-pi}
\end{figure}

After these selection criteria, we find
no candidate events in the modes \ModeEtapGG\ and \ModeEtapThreePZ, as
shown in Figure~\ref{fig:etap-eta-pi-pi}. However, in the mode
\ModeEtapPMZ, we find two good candidate events passing our selection
criteria as shown in Figure~\ref{fig:etap-eta-pi-pi}.
These two events have been looked at in detail and appear to be
good signal events. However, they are insufficient to allow
us to claim a positive signal, as no candidate events are observed 
in the modes \ModeEtapGG\ and \ModeEtapThreePZ, each providing higher
sensitivity than the decay chain \ModeEtapPMZ.

\subsection{The Decay $\Upsilon\to\gamma\eta^{\prime} ,\eta^{\prime}\to\gamma\rho^0$}
The reconstruction scheme for the decay chain \ModeEtapGR\ is slightly
different from those previously described. We first build $\rho^0$
candidates by forcing pairs of oppositely charged tracks to originate
from a common vertex. Next, we add a photon candidate 
(which we refer to as the ``soft shower'' having energy $E_s$ in
contrast with the high energy radiative photon)  
not associated with charged tracks, and having a lateral profile
consistent with being a photon, to build \etap\ candidates. To obtain
the maximum yield, we neither restrict the energy $E_s$ of the
photon nor the invariant mass of the $\rho^0$ candidate at this stage. 
A high energy photon is then added, ensuring that the soft shower and
high energy photon are distinct, to build the $\Upsilon$ candidate. The
$\Upsilon$ candidate is then constrained to the 4-momentum of the
initial $e^+e^-$ system and the candidate with the lowest
\chisqFourMom\ value is selected.

The candidate \etap\ invariant mass resolution is vastly improved due
to the mass-constraints on the candidate \pizero\ and $\eta$ mesons in
$\eta^{\prime} \to \eta \pi^+ \pi^-$ decays. In reconstruction of
\EtapGR, a significant improvement in candidate \etap\ invariant mass
resolution ($\approx 30\%$) as well as the energy resolution of the
soft shower is achieved by performing the 4-momentum constraint on
the $\Upsilon$ candidate.

Particle identification in the channel \EtapGR\ is achieved by
demanding the combined RICH and \dedx\ likelihood for the pion
hypothesis be greater than the combined likelihood for each of the
electron, kaon and proton hypotheses. Copiously produced QED processes
such as $e^+e^- \to \gamma \gamma e^+e^-$ are
suppressed by imposing an electron veto, requiring that
$|E/p-1.0|>0.05$, where $p$ is the measured momentum and $E$ is the
associated calorimeter energy of the charged track. QED events of the type 
$e^+e^- \to \gamma \gamma \mu^+\mu^-$ are suppressed by requiring that
neither track registers a hit five hadronic interaction lengths deep
into the muon detector system. Continuum background of the type 
$e^+e^- \to \gamma \gamma \rho^0$ is suppressed by demanding
$E_s>100$\mev. Finally, the event is ensured to be complete by
demanding \chisqFourMom\ $< 100$. 
The overall efficiency of the selection criteria for this mode is
$(40.1\pm2.1)\%$, including possible systematic uncertainties and
the statistical uncertainty of the Monte Carlo sample.

\begin{figure}
\includegraphics*[width=3.25in]{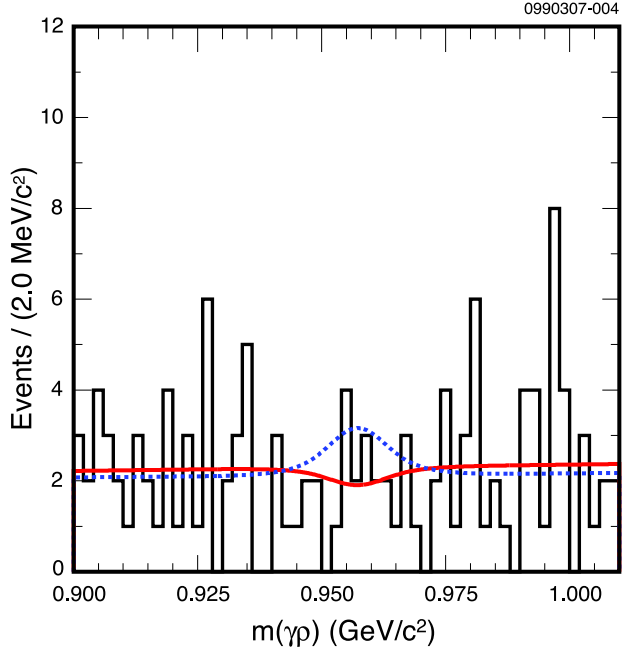}
\caption{Invariant mass distribution of $\gamma\rho^0$ candidates in $\Uos$
  data for the mode \ModeEtapGR\ overlaid with fits using a) floating area
  (solid red) yielding $-3.1\pm5.3$ events, and b) area fixed to
  8.6 events (dashed blue), corresponding to the upper limit at 90\% C.L.}
\label{fig:etap-gr}
\end{figure}

Although highly efficient, our selection criteria are not sufficient to
suppress the smooth continuum background from the reaction
$e^+e^- \to \gamma \gamma \rho^0$.
The candidate \EtapGR\ invariant mass distribution after our selection
criteria, shown in Figure~\ref{fig:etap-gr}, is fit to
a double Gaussian function over a floating polynomial background
function of order one. The parameters of the double Gaussian function
are fixed to the values obtained from a fit to signal Monte
Carlo and the area is left to float. The likelihood fit yields
$-3.1\pm5.3$ events, which is consistent with zero. In the absence of a
clear signal, we determine the upper limit yield
as we do in the case of \ModeEtaGG, and find an upper limit at 90\%
confidence level of 8.6
events.

\section{Systematic Uncertainties and Combined Upper Limits}\label{sec:sys-lim}
Since we do not have a signal in any of the modes, and since the
kinematic efficiency is near-maximal, statistical uncertainties
dominate over systematic uncertainties. 
By comparison of the expected yield of the QED process \qedThreePhoton\
with the calculated cross-section for this process, we estimate the
uncertainty on the trigger simulation for ``all neutral'' modes to be
4.5\%. For modes with only two charged tracks, we have studied the 
QED processes $e^+e^- \to \gamma \rho^0$ and $e^+e^- \to \gamma \phi$, 
and assign a 13\% uncertainty
on the efficiency due to possible trigger mismodeling. For events with
many charged tracks, we assign a systematic
uncertainty of 1\% as the relevant trigger lines are very well understood,
redundant, and
very efficient.
We assign 1\% uncertainty per track in charged track
reconstruction based upon CLEO studies~\cite{CLEOSystematics}
of low-multiplicity events, and
2.5\% systematic uncertainty per photon from mismodeling of
calorimeter response which translates to 5\% uncertainty per meson
(\pizero\ and $\eta$) decaying into \diphoton, 
again based upon CLEO studies~\cite{CLEOSystematics}.
The systematic uncertainty in $\SDeDx$ for two
tracks added in quadrature (as in $\ModeEtaPMZ$) was evaluated to be 4\%
by considering the efficiency difference of this requirement in
Monte Carlo and data samples of
\qedOmega. Consequently, we assign 4\% and 5.7\% uncertainty to the
reconstruction efficiencies of modes involving two and four charged
tracks, respectively, excepting \EtapGR\ where this requirement was not
imposed. For the mode \EtapGR, the systematic uncertainty in the
efficiency of analysis cuts, found to be 3.9\%, was evaluated by
comparing the efficiency difference in Monte Carlo and data by
studying the $\rho^0$ signal due to the QED processes. 
For the neural-net cut in the mode \ModeEtaGG, we studied the
efficiency in QED \qedThreePhoton\ simulated events and the real data
dominated by the same QED process for a wide range of neural-net
output values. We find a maximum difference of 7\% in these two
numbers, which we take as a conservative estimate of the
associated systematic uncertainty.
The systematic uncertainties for various $\eta$ and \etap\ decay modes
are listed in Table~\ref{tab:sys-errs}. 
These uncertainties were added in quadrature, along with the
statistical error due to the limited size of Monte Carlo samples, to
obtain the overall systematic uncertainties in the efficiencies.

\begin{small}
\begingroup
\squeezetable
\begin{table*}[!]
\begin{center}
\caption{\label{tab:sys-errs}Contributions to systematic
uncertainties in the efficiencies for \ModeEtap\ (upper half) and
\ModeEta\ (lower half). The uncertainties are expressed as relative
percentages and combined in quadrature.}
\begin{ruledtabular}
\begin{tabular}{lcccc}
Uncertainty source & ~~\EtapEtaGG & ~~\EtapEtaPMZ & ~~\EtapEtaThreePZ
& \EtapGR \\  

\hline
Trigger mismodeling    & 13   & 1   & 13    & 1   \\
Track reconstruction   & 2   & 4   & 2    & 2   \\
Calorimeter response   & 5   & 5   & 15   & 2.5 \\
Analysis cuts          & 4   & 5.7 & 4    & 3.9 \\
Monte Carlo statistics & 1.0 & 1.6 & 2.4  & 1.0 \\
\hline
Combined uncertainty   & 14.7 & 8.8 & 20.4 & 5.2 \\
\hline
& & & & \\
Uncertainty source &   $\EtaGG$ & $\EtaPMZ$ & $\EtaThreePZ$ & \\
\hline
Trigger mismodeling    & 4.5   & 13   & 4.5    & \\
Track reconstruction   & -   & 2   & -    & \\
Calorimeter response   & 5   & 5   & 15   & \\
Analysis cuts          & 7   & 4   & -    & \\
Monte Carlo statistics & 1.3 & 1.2 & 1.7  & \\
\hline
Combined uncertainty   & 9.8 & 15.2 & 16.0 & \\
\end{tabular}
\end{ruledtabular}
\end{center}
\end{table*}
\endgroup
\end{small}

\begin{small}
\begingroup
\squeezetable
\begin{table*}[ht]
\begin{center}
\caption{Results of the search for \ModeEtap\ and \ModeEta. Results include
statistical and systematic uncertainties, as described in the
text. The combined limit is obtained after including the systematic
uncertainties.}
\label{tab:limits}
\begin{ruledtabular}
\begin{tabular}{lcccc}
& ~~\EtapEtaGG & ~~\EtapEtaPMZ & ~~\EtapEtaThreePZ & ~~\EtapGR \\
\hline

Observed events & 0 & 2 & 0 & $-3.1\pm5.3$ \\

$\mathcal{B}_{\eta^{\prime},i}\%$ & 
$17.5\pm0.6$ & $10.0\pm0.4$ & $14.4\pm0.5$ & $29.5\pm1.0$ \\

Reconstruction efficiency (\%) & 
$35.2\pm5.2$ & $24.5\pm2.2$ & $14.4\pm2.9$ & $40.1\pm2.1$ \\ 

$\mathcal{B}(\Uos\to\gamma\eta^{\prime})(90\%~\text{C.L.})$\footnotemark[1] &
\ulbr{1.8} & \ulbr{10.3} & \ulbr{5.2} & \ulbr{3.4} \\

$\mathcal{B}(\Uos\to\gamma\eta^{\prime})(90\%~\text{C.L.})$\footnotemark[2] &
\ulbr{1.9} & \ulbr{10.4} & \ulbr{5.8} & \ulbr{3.4} \\ 

\hline

\multicolumn{2}{l}{Combined limit on 
$\mathcal{B}(\Uos\to\gamma\eta^{\prime})$} &
\multicolumn{2}{c}{\ulbr{1.9}} \\

\hline

& & & & \\

& $\EtaGG$ & $\EtaPMZ$ & $\EtaThreePZ$ & \\

\hline

Observed events & $-2.3\pm8.7$ & 0 & 0 & \\

$\mathcal{B}_{\eta,i}\%$ & $39.4\pm0.3$ & $22.6\pm0.4$ & $32.5\pm0.3$ & \\

Reconstruction efficiency (\%) & $23.8\pm2.4$ & $28.5\pm2.9$ &
$11.8\pm1.9$ & \\

$\mathcal{B}(\Uos\to\gamma\eta)(90\%~\text{C.L.})$\footnotemark[1]
& \ulbr{7.3} & \ulbr{1.7} & \ulbr{2.8} & \\

$\mathcal{B}(\Uos\to\gamma\eta)(90\%~\text{C.L.})$\footnotemark[2] & 
\ulbr{7.4} & \ulbr{1.8} & \ulbr{2.9} & \\
\hline

Combined limit on $\mathcal{B}(\Uos\to\gamma\eta)$ &
\multicolumn{3}{c}{\ulbr{1.0}} &\\

\end{tabular}
\end{ruledtabular}
\footnotetext[1]{excluding systematic uncertainties}
\footnotetext[2]{including systematic uncertainties}
\end{center}
\end{table*}
\endgroup
\end{small}

\begin{figure*}[!]
\mbox{\includegraphics*[width=6.3in]{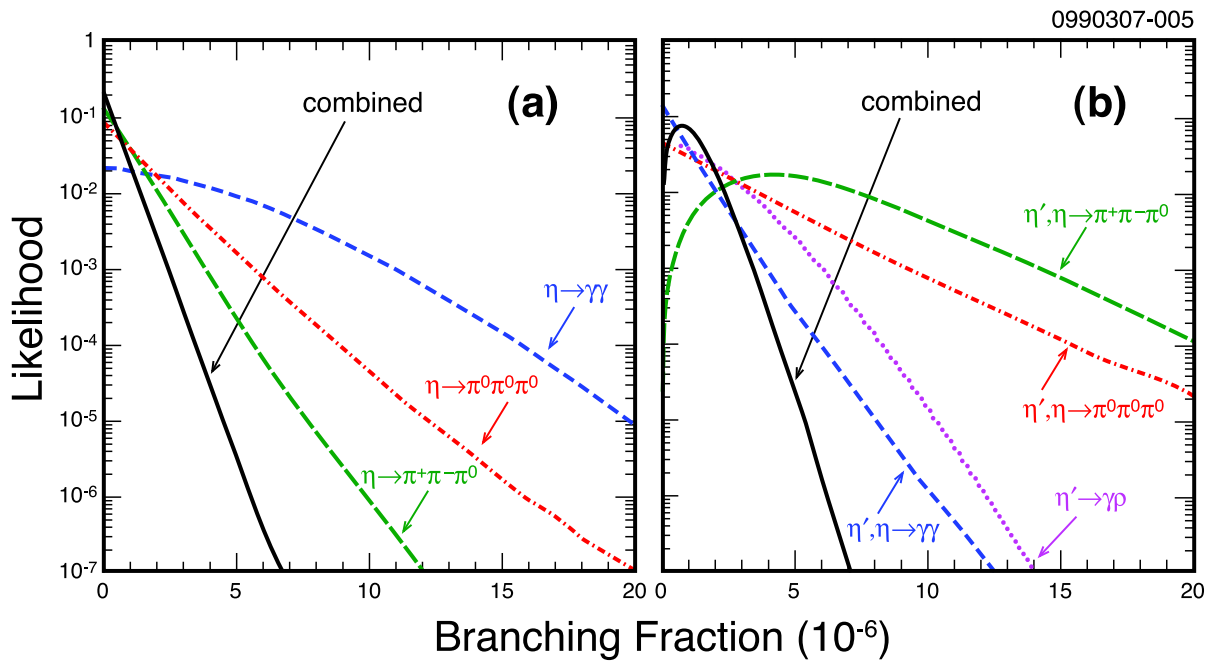}}
\caption{Likelihood distributions as a function of branching fraction
for the decay mode \ModeEta\ (left) and \ModeEtap\ (right). All
distributions are smeared by respective systematic uncertainties and
normalized to the same area. The solid black curve denotes the combined
likelihood distribution.} 
\label{fig:limit-plots}
\end{figure*}
The systematic uncertainties in efficiencies, uncertainties in the
product branching ratios, and the statistical uncertainty in the number
of $\Uos$ decays, \Nups, are incorporated~\cite{Cousins}
by a ``toy'' Monte Carlo procedure to obtain smeared likelihood
distributions for the branching fraction in each mode,
$\mathcal{B}(\Uos\to\gamma\text{P}) = N_{\text{P}}/(\sensitvt)$, 
where $\text{P} = \eta , \eta^{\prime}$, and $\epsilon_i$ and
$\mathcal{B}_{\text{P},i}$ denote the efficiency and branching
fractions of the $i$th mode. 
To obtain the smeared likelihood distribution $\mathcal{L}_{\text{P},i}$, the
experiment is performed multiple times, randomly selecting
$N_{\text{P}}$  from the likelihood function appropriate for each
mode\footnote{For modes with zero or few observed events, the
  appropriate likelihood function is generated from Poisson statistics.
  For the background limited modes $\EtaGG$ and \EtapGR, we 
  already have the likelihood function which we used in calculating
  the upper limit of the observed number of events at 90\% CL.} and then
dividing by the sensitivity factor $\sensitvt$, where each term is picked
from a Gaussian distribution about their mean values with the
appropriate standard deviation. 

The combined likelihood distribution for
$\mathcal{B}(\Uos\to\gamma\text{P})$ is derived as
$\mathcal{L}_{\text{P}} = \prod_{i}{\mathcal{L}_{\text{P},i}}$ which
is summed up to 90\% of the area in the physically allowed region to obtain
the upper limit branching fraction for $\Uos\to\gamma\text{P}$.
From the constituent $\mathcal{L}_{\text{P},i}$ and the combined
$\mathcal{L}_{\text{P}}$ as shown in 
Figure~\ref{fig:limit-plots}, 
we obtain upper limits on
$\mathcal{B}(\Uos\to\gamma\eta)$ of \br{7.4}, \br{1.8}, \br{2.9}, 
and \br{1.0} for $\eta$ decaying into \diphoton,
$\pi^+\pi^-\pi^0$, $\pi^0\pi^0\pi^0$, and all three combined,
respectively. We obtain upper limits for
$\mathcal{B}(\Uos\to\gamma\eta^{\prime})$ 
of \br{1.9}, \br{10.4}, \br{5.8}, and \br{3.4} for $\eta$
decaying into \diphoton, $\pi^+\pi^-\pi^0$, $\pi^0\pi^0\pi^0$,
and \EtapGR, respectively. The combined upper limit for \BrModeEtap\ is
$1.9\times10^{-6}$, a value larger than one of the sub-modes
(\ModeEtapGG), due to the two candidate events in \ModeEtapPMZ.
The numbers of observed events, detection efficiencies and upper
limits are listed in Table~\ref{tab:limits}.

\section{Summary and Conclusion}\label{sec: summary}
We report on a new search for the radiative decay of $\Uos$ to the 
pseudoscalar mesons $\eta$ and \etap\ in $21.2\times10^{6}$ $\Uos$ 
decays collected with the CLEO~III detector. The $\eta$ meson was reconstructed
in the three modes $\EtaGG$, $\EtaPMZ$ or $\EtaThreePZ$.
The \etap\ meson was reconstructed either in the mode \EtapGR\ or 
$\eta^{\prime} \to \pi^{+} \pi^{-} \eta$ with $\eta$ decaying through any
of the above three modes. All these modes except for \EtapGR\ had earlier
been investigated in CLEO~II data amounting to 
\Nups $ = 1.45\times 10^6$ $\Uos$ mesons and resulted in
previous upper limits \BrModeEtap\ $< 1.6 \times 10^{-5}$ and 
\BrModeEta\ $< 2.1 \times 10^{-5}$  
at 90\% C.L. These limits were already smaller than 
the naive predictions based upon the scaling of the decay rate for the
corresponding \Jpsi\ radiative decay mode by the factor 
$(q_{b}m_{c}/q_{c}m_{b})^{2}$,
and also the model of K\"orner \etal~\cite{KKKS}, 
whose perturbative QCD approach predictions
for $\mathcal{B}(J/\psi\to\gamma X)$ where $X = \eta,
\eta^{\prime}, f_2$ as well as $\mathcal{B}(\Uos\to\gamma f_2)$ agree
with experimental results.

With a CLEO~III data sample 14.6 times as large as the CLEO~II data
sample, we find no convincing signal in any of the modes. Based purely
upon the luminosities, we would expect the new upper limits to be scaled 
down by a factor of between 14.6 (in background-free modes) and
$\sqrt{14.6}$ in background dominated modes if the two CLEO detectors 
(CLEO~II and CLEO~III) offered similar particle detection efficiencies. 
In the search for \ModeEta\ we find no hint of 
a signal, and manage to reduce the limit by an even larger factor. In
the search for \ModeEtap, however, we find two clean candidate events
in the channel \ModeEtapPMZ, which, though we
cannot claim them as signal, do indicate the possibility that we are
close to the sensitivity necessary to obtain a positive result. 
Because of these two events, our combined limit for \ModeEtap\ is not 
reduced by as large a factor as the luminosity ratio, and in fact is 
looser than that which would be obtained if we analyzed 
one sub-mode (\ModeEtapGG) alone.
In this analysis we found upper limits which we
report at 90\% confidence level as
\begin{displaymath}
\mathcal{B}(\Uos \to \gamma \eta) < 1.0 \times 10^{-6},
\end{displaymath}
\begin{displaymath}
\mathcal{B}(\Uos \to \gamma \eta^{\prime}) < 1.9 \times 10^{-6}.
\end{displaymath}

Our results are sensitive enough to test the appropriateness of the
pseudoscalar mixing approach as pursued by Chao~\cite{KTChao}, where
mixing angles among various pseudoscalars including \etab\ are
calculated. Then, using a calculation for the M1 transition
$\Upsilon\to\gamma\eta_{b}$, he predicts 
$\mathcal{B}(\Uos \to \gamma \eta) = 1\times 10^{-6}$ and
$\mathcal{B}(\Uos \to \gamma \eta^{\prime}) = 6\times 10^{-5}$. Our
limit for \ModeEtap\ is significantly smaller than Chao's prediction
and does not support his approach.

The sensitivity challenge posed by both the extended vector dominance
model and the higher twist approach of Ma are beyond our reach. 
In extended VDM, Intemann predicts 
$1.3\times 10^{-7} < \mathcal{B}(\Uos\to\gamma\eta) < 6.3\times 10^{-7}$ 
and 
$5.3\times 10^{-7} < \mathcal{B}(\Uos\to\gamma\eta^{\prime}) <
2.5\times 10^{-6}$, where the two limits are determined by having
either destructive or constructive interference, respectively, between
the terms involving $\Uos$ and $\Utwos$. Even if it is determined that 
the amplitudes are added constructively, our limit remains higher than
the VDM prediction for $\Uos\to\gamma\eta$. 

Ma's prediction of 
$\mathcal{B}(\Uos\to\gamma\eta^{\prime}) \approx 1.7\times 10^{-6}$ is
consistent with our result. However, his prediction for
$\mathcal{B}(\Uos\to\gamma\eta) \approx 3.3\times 10^{-7}$ is a factor of
$\sim 3$ smaller than our limit.

We gratefully acknowledge the effort of the CESR staff 
in providing us with excellent luminosity and running conditions. 
D.~Cronin-Hennessy and A.~Ryd thank the A.P.~Sloan Foundation. 
This work was supported by the National Science Foundation,
the U.S. Department of Energy, and 
the Natural Sciences and Engineering Research Council of Canada.

\end{document}